\input{aipcheck}

\documentclass[
    ,final            
  ]
  {aipproc}

\layoutstyle{6x9}

\newcommand{\ee}{\end{equation}} 
\newcommand{\be}{\begin{equation}} 
\newcommand{\ec}{\end{center}} 
\newcommand{\bc}{\begin{center}} 
\newcommand{\eea}{\end{eqnarray}} 
\newcommand{\bea}{\begin{eqnarray}} 
\newcommand{\bd}{\begin{description}} 
\newcommand{\ed}{\end{description}} 
\newcommand{\bi}{\begin{itemize}} 
\newcommand{\ei}{\end{itemize}} 
 
\def\vx{\vec{\mbox{$x$}}}
\def\vy{\vec{\mbox{$y$}}}
\def\vk{\vec{\mbox{$k$}}}

\def\Tr{{\rm Tr}}

\def\spose#1{\hbox to 0pt{#1\hss}}
\def\ltapprox{\mathrel{\spose{\lower 3pt\hbox{$\mathchar"218$}}
 \raise 2.0pt\hbox{$\mathchar"13C$}}}
\def\gtapprox{\mathrel{\spose{\lower 3pt\hbox{$\mathchar"218$}}
 \raise 2.0pt\hbox{$\mathchar"13E$}}}

\begin{document}

\title{Lattice Results in Coulomb Gauge}


\author{Attilio Cucchieri}{
  address={Instituto de F\'\i sica de S\~ao Carlos, Universidade de S\~ao Paulo, 
Caixa Postal 369, \\ 13560-970 S\~ao Carlos, SP, Brazil}
}

\begin{abstract}
We discuss recent numerical results obtained for gluon and ghost
propagators in lattice Coulomb gauge and the status of the so-called
Gribov-Zwanziger confinement scenario in this gauge.
Particular emphasis will be given to the eigenvalue spectrum
of the Faddeev-Popov matrix.
\end{abstract}

\maketitle


\section{Introduction}

In recent years several groups have studied confinement of quarks and
gluons using lattice simulations in Coulomb gauge
\cite{Cucchieri:2000gu}--\cite{Grady}.
This gauge has several advantages, even though it breaks the Lorentz symmetry
explicitly. Indeed, it is a physical gauge \cite{TDL}, i.e.\ there are no
unphysical degrees of freedom. Also, a nice confinement scenario
\cite{Gribov:1977wm}--\cite{reviews}
is available for this gauge, based on Gribov's classical work
\cite{Gribov:1977wm}. Finally, Coulomb gauge
is well-suited for the Hamiltonian
approach and the study of hadron physics by variational methods
\cite{hamiltonian}.

Here we review lattice numerical studies in Coulomb gauge. We divide these
studies in two periods. In the first period --- called here the {\em classical
era} --- the studies focused on the infrared (IR) behavior of
propagators (gluon and ghost) \cite{Cucchieri:2000gu}--\cite{Langfeld:2004qs}
and on the long-distance behavior of the color-Coulomb potential
\cite{Cucchieri:2002su}--\cite{Nakagawa:2006fk}.
On the other hand, in the second period --- the {\em modern era} --- 
the eigenvalue spectrum of
the Faddeev-Popov (FP) operator became the main subject of investigation
\cite{Greensite:2004ur,Nakagawa:2006at}.


\section{Coulomb gauge at the classical level}

In a classical Yang-Mills theory \cite{TDL}, Gauss's law is written as
$\, \left( D_{i}\, E_{i} \right)^{a}(\vx\mbox{,}\, t) \,=\,
\rho^{a}_{qu}(\vx\mbox{,}\, t)\,$,
where $D_{i}$ is the gauge-covariant derivative,
$\,E^{a}_{i}(\vx\mbox{,} t)$ is the color-electric field
and $\rho^{a}_{qu}(\vx\mbox{,}\, t)$ is the quark color-charge density.
(Here the sum over repeated indices is always understood.)
In Coulomb gauge, i.e.\ $(\partial_{i}\, A_{i}\,)^a(\vx\mbox{,}\, t)\,=\,0$,
the color-electric field can be decomposed into its transverse
and longitudinal parts using $\, E^{a}_{i}(\vx\mbox{,}\, t) \,\equiv\,
( E^{tr} )^{a}_{i}(\vx\mbox{,}\, t) \,-\, \partial_{i} \phi^{a}(\vx\mbox{,}\, t)\,$,
where $\phi^{a}(\vx\mbox{,}\, t)$ is the so-called {\em color-Coulomb potential}
and $\, ( \partial_{i} \, E_{i}^{tr} )^{a}(\vx\mbox{,}\, t) \,=\,0\,$.
Then, Gauss's law becomes $\, \left(\, {\cal M} \,\phi^{a}\,\right)(\vx\mbox{,}\, t)\,=\,
\rho^{a}(\vx\mbox{,}\, t)\,$,
where $\,{\cal M}\equiv - D_{i}\,\partial_{i}\,$ is the 3-dimensional
FP operator and $\, \rho^{a}(\vx\mbox{,}\, t) \, =\,
\rho^{a}_{q}(\vx\mbox{,}\, t) \,-\, f^{a b c}\,A^{b}_{i}(\vx\mbox{,}\, t)\,
\left(\,E^{tr}_{i}\,\right)^{c}(\vx\mbox{,}\, t) \,$
is the total color-charge density.
It follows that the color-Coulomb potential
$\phi^{a}(\vx\mbox{,}\, t)$ can be expressed by means of the instantaneous and
non-local operator $( {\cal M}^{-1} )^{a b}(\vx, \vy; t)$, namely
$\, \phi^{a}(\vx, t) \,=\, \left( {\cal M}^{-1} \rho \right)^{a}(\vx, t) \,
=\, \int\, d^{3}y \, \left( {\cal M}^{-1} \right)^{a b}(\vx, \vy; t)
\, \rho^{b}(\vy, t) \,$.
At the same time, the classical Hamiltonian $\, {\cal H} \,=\, 
\int d^3x \; (\, E_{i}^2 \, + \, B_{i}^2 \, ) / 2 \, $
can be written as $\,
{\cal H} \,=\, {\cal H}_{Coul}
  \,+\, \int d^3x \; \left[ (E^{tr}_{i})^2 \, + \, B_{i}^2 \right] /2 \,$.
Here
\be
{\cal H}_{Coul} \,=\, \frac{1}{2} \,
\int d^3x \;\; (\partial_{i}\, \phi)^2 \, = \, 
\frac{1}{2} \,
\int d^3x \int d^3y \; \rho^a(\vx) \, {\cal V}^{ab}(\vx,\vy) \, \rho^b(\vy)
\;\mbox{,}
\vspace{-2mm}
\ee
$ \,{\cal V}^{ab}(\vx,\vy) \,=\, \left[ {\cal M}^{-1} (-\Delta)
{\cal M}^{-1} \right]^{ab}(\vx,\vy) $
is the color-Coulomb-potential energy functional and we indicate with
$\Delta$ the usual Laplacian.

Clearly, the static color-Coulomb potential $\phi^{a}(\vx\mbox{,}\, t)$ is
closely related to the 3d FP (equal-time) ghost propagator
$\, G(\vx - \vy, t) \,\delta^{a, b} \,\equiv\,
    \langle\,\left( {\cal M}^{-1} \right)^{a b}(\vx, \vy; t)\,\rangle \,$.
In particular, if we consider the Fourier
transform ${\widetilde G}(\vk, t)$
we obtain
\be
\phi^{a}(\vx, t) \,\approx\,
\frac{1}{(\,2\,\pi\,)^{3}}\,\int\,d^{3}k\, \int\, d^{3}y \;\,
{\widetilde G}(\vk, t) \; \exp{\left[\, i\, k_{j}
      \left( \, x_j - y_j\,\right)\right]}
\,\rho^{a}(\vy\mbox{,}\, t)
\;\mbox{.}
\ee
Thus, if the ghost propagator has a $k^{-4} = |\vk|^{- 4}$ singularity at small
momenta we get, in the limit of large separation $x=|\vx|$,
a linearly rising potential, i.e.\
$\, \phi^{a}(\vx, t) \,\sim\,x \,$.


\section{The Gribov-Zwanziger confinement scenario}

A non-perturbative investigation of QCD in the IR limit
is necessary in order to get an understanding of
color confinement.
Of course, in developing non-perturbative techniques,
one has to deal with the redundant gauge degrees
of freedom of the theory. The gauge-fixing technique
developed by Faddeev and Popov assumed
that one could find a gauge-fixing condition that
uniquely determines a gauge field on each gauge orbit.
However, in Ref. \cite{Gribov:1977wm} Gribov
showed that the Coulomb and the Landau
gauge conditions do not fix the gauge fields uniquely,
namely there exist gauge-related
field configurations that satisfy the gauge condition
({\em Gribov copies}) \cite{Gribov:1977wm,reviews}.

In order to get rid of the problem of spurious gauge copies, Gribov
proposed the use of additional gauge conditions. In
particular, for Coulomb gauge, he proposed the restriction
of the physical configuration space (on each time-slice $t$) to the region
$\, \Omega_{t} \equiv \{\, A: \, (\partial_{i}\, A_{i}\,)^a(\vx\mbox{,}\, t)\,=\,0 
      \mbox{,} \; {\cal M}^{a b}(\vx, \vy; t) \,\geq\,0\, \} \,$.
Thus, inside the region $\Omega_t$, the FP operator has
no negative eigenvalues. This region is 
delimited by the so-called {\em first Gribov horizon} $\,\partial \Omega_t$, where
the smallest non-trivial eigenvalue
of the FP operator ${\cal M}^{a b}(\vx, \vy; t)$ is zero.
On the lattice, given a thermalized lattice configuration
$\,\{U (x)\}\,$, a configuration belonging to the region
$\Omega_t \,$ can be obtained by finding a gauge
transformation $\,\{g (x)\}\,$ that brings the
functional\footnote{In this review we do not discuss results related to the (possible)
spontaneous symmetry breaking of the residual gauge freedom $g(t)$
\cite{Greensite:2004ke,Nakamura:2005ux,Grady}
and to the so-called $\lambda$ gauge \cite{lambda}, which
interpolates between the Landau gauge ($\lambda = 1$)
and a Coulomb-gauge like condition ($\lambda \to 0$).}
$\, {\cal E}_{\mbox{hor}, U}[ g ]\,=\,- \,\sum_{i = 1}^{3}\,
  \sum_{\vx, t}\,\Tr\,
\left[\,g(\vx, t)\, U_{i}(\vx, t)\, g^{\dagger}(\vx + a \, e_i, t) \, \right]
\,$ to a local minimum.
Recall that, in the $SU(N_c)$ case, both the link variables
$\,U_{\mu}(x)\,$ and the gauge transformation matrices
$\,g(x)\,$ are elements of the $\,SU(N_c)\,$ group (in
the fundamental $\,N_c \times N_c \,$ representation).

The additional gauge condition added by Gribov is not significant
for the high-frequency vacuum fluctuations, i.e.\ for the perturbative regime,
but it suppresses the low-frequency fluctuations, modifying the
(non-perturbative) IR regime \cite{Gribov:1977wm,Zwanziger}.
In particular, 
one can show that, when the functional integration is restricted
to the region $\Omega_t$, then (on each time slice) the  
ghost propagator $G(\vk, t)$ is IR enhanced. On the other hand,
the transverse gluon propagator $D^{tr}(\vk, t)$
may go to zero in the IR limit, implying a maximal violation
of reflection positivity.
The latter result may be viewed as an indication of gluon confinement
\cite{Alkofer:2000wg}.
Analytic results for the IR behavior of propagators and vertices
using Dyson-Schwinger equations have been
presented in Ref.\ \cite{Schleifenbaum:2006bq}.

At the same time, the $44$-component of the gluon propagator can be written
\cite{Zwanziger2} as $ \, D_{44}(\vx-\vy,t) \, = \,
V_{Coul}(\vx-\vy) \, \delta(t) + P(\vx-\vy,t)\,$,
where $\, V_{Coul}(\vx-\vy) \,
\delta^{ab} = \langle {\cal V}^{ab}(\vx,\vy) \rangle \,$
is anti-screening and should yield a linearly rising potential,
while $P(\vx-\vy,t)$ is the vacuum-polarization term, i.e.\ it is
responsible for screening and for the breaking of the string between
color sources. One can show that
these three quantities [e.g.\ $D_{44}(\vx-\vy,t)$, $V_{Coul}(\vx-\vy)$
and $P(\vx-\vy,t)$] are renormalization-group invariant
\cite{Zwanziger2,Cucchieri:2001zb}. One can also
define the running coupling constant
\be
g^2_{Coul}(\vk) \; = \; \frac{11 N_c - 2 N_f}{12 N_c}
 \;\; k^2 \; V_{Coul}(\vk) 
\;\mbox{.}
\label{eq:g2}
\ee
Clearly, if the color-Coulomb potential $V_{Coul}(x)$ is linearly rising at
large separation $x$, then in the IR limit we find $V_{Coul}(\vk) \sim 1/k^{4}$
and $g^2_{Coul}(\vk) \sim 1/k^{2}$. 
Also, it has been shown \cite{Zwanziger:2002sh}
that the Coulomb energy of static sources is an
upper bound for the static inter-quark potential $V(\vx)$,
i.e.\ if at large $x$ one has
$V_{Coul}(\vx) = \sigma_{Coul} \, x$ and
$V(\vx) = \sigma\, x$ then we find $(N_c^2 - 1) \sigma_{Coul} / (2 N_c) \geq \sigma$.
Analytic results for the long distance behavior of $V_{Coul}(\vx)$ have been
presented in Ref.\ \cite{Zwanziger:2003de}.

Summarizing \cite{Zwanziger:2006sc},
in the Gribov-Zwanziger confinement scenario (in Coulomb gauge),
the long-range force, responsible for color confinement, is carried by an
instantaneous static color-Coulomb field. In particular, the
linearly rising potential is related to the IR divergence of the
ghost-propagator (at equal time). At the same time, the
propagator of three-dimensionally transverse (would-be physical) gluons
is IR suppressed and the gluons are absent from the spectrum.


\section{The classical era: results}

The analytic predictions described above for the gluon
propagators $D^{tr}(\vk)$ and $D_{44}(\vk)$ have been
verified for the $SU(2)$ group in Refs.\ 
\cite{Cucchieri:2000gu}--\cite{Langfeld:2004qs}.
In particular, from
Fig.\ 1 of Ref.\ \cite{Cucchieri:2000gu} it is evident that, in the IR limit,
the transverse propagator is suppressed, while $D_{44}(\vk)$
blows up. Moreover, in the infinite-volume limit, it has been
found \cite{Cucchieri:2000gu,Cucchieri}
that $D^{tr}(\vk)$ is well described a Gribov-like propagator
characterized by a pair of purely imaginary poles $m^2 = \pm i y$.
Numerically, at $\beta = 2.2$ and in the infinite-volume limit, one finds
$\,y = 0.33 \pm 0.14$ GeV$^2$.
As for the ghost propagator $G(\vk,t)$, it has been studied up to now only in
Ref.\ \cite{Langfeld:2004qs}. There, it is shown that $G(\vk,t)$ has indeed an IR
divergence stronger than $1/k^2$.
At the same time, the running coupling $g^2_{Coul}(\vk)$, defined in Eq.\ (\ref{eq:g2})
above, seems to be consistent \cite{Cucchieri:2002su,Langfeld:2004qs}
with an IR behavior of the type $1/k^{2}$. These analyses have also
obtained $\sigma_{Coul} \approx \sigma$.

In Ref.\ \cite{Greensite:2003xf},
the color-Coulomb potential $V_{Coul}(\vx)$ has been evaluated [for the $SU(2)$
group]\footnote{Similar results were obtained for the $SU(3)$ group in Refs.\
\cite{Nakamura:2005ux,Nakagawa:2006fk}.}
as a function of the separation $\,x\,$, using correlators of two time-like
Wilson lines of length 1 (in lattice units). It was found
that $V_{Coul}(\vx)$ increases linearly with $x$, in agreement with
the $1/k^{2}$ behavior for $g^2_{Coul}(\vk)$ obtained in Refs.\
\cite{Cucchieri:2002su,Langfeld:2004qs}. However, in this case
the estimate for the Coulomb string tension was
$\sigma_{Coul} \approx 2-3 \, \sigma$.
Moreover, if one removes the so-called center vortices \cite{Greensite:2003xf},
then the color-Coulomb potential $V_{Coul}(\vx)$ goes to a constant at large $x$ and
$\sigma_{Coul} = 0$. This suggests a strong relation between these
center vortices and the (Coulomb) confinement mechanism. Note that similar
effects have been observed in the gluon and in the ghost propagators
in Landau gauge \cite{Gattnar:2004bf} after removing the center vortices.

It is also interesting that, when the temperature is turned on
\cite{Nakamura:2005ux,Nakagawa:2006fk},
the color-Coulomb potential $V_{Coul}(\vx)$ is not screened and
it is still a linearly rising function of $x$. Moreover, the
Coulomb string tension $\sigma_{Coul}$ shows a magnetic-like behavior
\cite{Nakagawa:2006fk}, i.e.\ $\sigma_{Coul}^{1/2} \sim g^2(T) T$.
This implies that the Coulomb string tension cannot be used as an order parameter
for confinement. This conclusion can be understood by
observing that the temperature is defined by compactifying the time direction
and that the Coulomb gauge is defined on the subspace orthogonal 
to the time direction. 
Thus, there is no reason for the system in Coulomb gauge to be sensitive to
the deconfining transition.


\section{The Modern Era}

In Ref.\ \cite{Greensite:2004ur} the authors evaluate the gauge-field
excitation energy ${\cal E}$ (above the ground state energy) of
a single (point-like) static color charge in Coulomb gauge. Considering
that long-range effects should be related to the non-local-interaction
term ${\cal H}_{Coul}$, one finds
\be
{\cal E} \, \propto \, {\cal V}^{aa}(x,x) \,=\,
 \left[ {\cal M}^{-1} (-\Delta) {\cal M}^{-1} \right]^{aa}(x,x)
\; \mbox{.}
\label{eq:calE}
\ee
A necessary condition for confinement is that ${\cal E}$ should diverge in
the infinite-volume limit, due to IR effects. (Ultraviolet divergences
are regulated by the lattice cut-off.)
For the (Coulomb) FP matrix
${\cal M}^{ab} = - \delta^{ab} \Delta - f^{acb} A^c_{\mu} \partial_{\mu}\,$
one can consider (inside the Gribov region $\Omega_t$)
the eigenvalues $\lambda > 0$
and the corresponding eigenfunctions $\Phi^a_{\lambda,x}$.
Then, Eq.\ (\ref{eq:calE}) can be written as $\;
{\cal E} \,\propto \, \langle \sum_{\lambda} F_{\lambda} / \lambda^2 \rangle \,$
with
$\, F_{\lambda} \,=\, 
V_s^{-1} \sum_{xy} (\Phi^a_{\lambda,x})^* \, (- \Delta)_{x,y}\,
      (\Phi^a_{\lambda,y}) \,$.
(Here, $V_s$ is the 3d spatial volume of the lattice.)
For a sufficiently large volume, the sums can be approximated by integrals
and $\, {\cal E} \,\propto \, \langle \int_{\lambda_{min}}^{\lambda_{max}} d\lambda \,
\rho(\lambda) F(\lambda) / \lambda^2 \rangle \, $,
with $\int d\lambda \, \rho(\lambda) = 1$.
In the infinite-volume limit, the volume of the Gribov region gets
concentrated near the Gribov horizon \cite{Greensite:2004ur,Zwanziger:2002sh,horizon},
i.e.\ $\lambda_{min} \to 0$. In the same limit, the gauge-field
excitation energy ${\cal E}$ blows up if
\be
\lim_{\lambda \to 0} \frac{\rho(\lambda) F(\lambda)}{\lambda} > 0 
\;\mbox{.}
\label{eq:conf}
\ee
Thus, a necessary condition for
confinement is the enhancement of $\rho(\lambda) F(\lambda)$ at small momenta.

In Appendix A of Ref.\ \cite{Greensite:2004ur}, an interesting analysis based on 
a random-matrix model shows that, for small eigenvalues,
one should have $\rho(\lambda) = c \lambda^{\alpha}$
if the eigenvalues $\lambda$ scale as $ V_s^{-1/(1+\alpha)}$
if the volume is increased.
Numerically they find $\alpha = 0.25 (5)$, implying
$\lambda_{min} \sim 1/L^{2.4}$.
At the same time, they obtain
$ F(\lambda_{min}) \sim 1/L \sim \lambda_{min}^{0.38} \,$ and
the confinement criterion (\ref{eq:conf}) is clearly fulfilled
\cite{Greensite:2004ur,Nakagawa:2006at}.
A similar result is obtained when considering the so-called
``vortex-only'' configurations \cite{Greensite:2004ur}.
On the other hand, after removing the center vortices, one
recovers a Laplacian-like eigenvalue spectrum for the FP
operator ${\cal M}^{ab}_{xy}$ with
$\lim_{\lambda \to 0} \rho(\lambda) F(\lambda) / \lambda = 0$.
Thus, in agreement with the findings reported in the previous
Section, the enhancement of
$\rho(\lambda) F(\lambda)$ at small eigenvalues $\lambda$
and the confinement mechanism in Coulomb gauge seem to be
strictly related to the properties of the
center-vortex configurations.
One can also show \cite{Greensite:2004ur} that center-vortex
configurations are (infinitely many) distinguished points on the Gribov
horizon. The relation between these configurations and the
Gribov-Zwanziger scenario in Coulomb gauge is then clear
if, in the infinite-volume limit, the center-vortex
configurations are sufficiently dense on the Gribov
horizon.


\section{An open question for the future}

In order to understand fully the Gribov-Zwanziger confinement scenario
one should consider a
generic gauge condition ${\cal F}[A]=0$, imposed by
minimizing a functional $E[U]$. Then, from the second variation of $E[U]$,
we can always define the FP matrix ${\cal M}^{ab}_{xy}$.
Clearly, when we are at a (local) minimum of $E[U]$,
the (non-trivial) eigenvalues of ${\cal M}^{ab}_{xy}$ are positive and
we can define a Gribov region $\Omega$ and the
first Gribov horizon $\partial \Omega$.
Moreover, since the configuration space has a very large
dimensionality, entropy should favor (in the limit of large volumes)
configurations near the Gribov horizon
\cite{Greensite:2004ur,Zwanziger:2002sh,horizon}, i.e.\ $\lambda_{min}$
should go to zero in the same limit.
This is indeed the case in 3d \cite{Cucchieri:2006tf}
and 4d Landau gauge \cite{Sternbeck:2005vs},
in 4d Coulomb gauge \cite{Greensite:2004ur}
and in 4d Maximally Abelian gauge (MAG) \cite{Mendes:2006kc}.
Since the FP matrix develops a null eigenvalue
at the Gribov horizon $\,\partial \Omega \,$,
we should also expect the
corresponding ghost propagator $\,G(k)\,$
to blow up at small momenta in the infinite-volume
limit. This result should in turn introduce a long-range effect
in the theory, being probably related to the color-confinement mechanism. Indeed, we
know from several numerical studies that
the ghost propagator $G(k)$ is IR enhanced
in 3d \cite{Cucchieri:2006tf} and 4d Landau gauge
\cite{Gattnar:2004bf,ghost,Oliveira:2006zg}
and in 4d Coulomb gauge \cite{Langfeld:2004qs}.
On the other hand, recent numerical results in MAG \cite{Mendes:2006kc}
suggest an IR finite $G(k)$.
Thus, the line of thinking reported above cannot be completely correct.
Of course, one
does not expect the ghost propagator to be particularly important
for confinement in MAG, since in this case the accepted scenario
is that confinement is related to Abelian dominance and (therefore)
to the IR behavior of the diagonal gluon propagator \cite{Mendes:2006kc,abelian}.
In any case, we should try to answer the following question:
{\em what makes the ghost propagator IR enhanced in Coulomb and in
Landau gauge but IR finite in MAG?}

A possible solution comes from the observation
that in Landau \cite{Cucchieri:2006tf,Sternbeck:2005vs}
and in Coulomb gauge \cite{Greensite:2004ur}
$ \lambda_{min} \sim 1 / L^{2+\alpha} $ with $ \alpha > 0 $,
i.e.\ it goes to zero faster than in the case of the Laplacian.
On the contrary, in MAG \cite{Mendes:2006kc} one has
$ \lambda_{min} \sim 1 / L^{2-\alpha} $ (with $ \alpha > 0$),
i.e.\ it goes to zero more slowly than for the Laplacian.
This (unproven) hypothesis seems to be supported by the following
observation. Using the same notation introduced in the previous section and
in the limit of a large volume, we can write the ghost propagator $G(k)$ as
\be
G(k) \, = \,\int_{\lambda_{\min}}^{\lambda_{max}} d\lambda\;\frac{\rho(\lambda)
f_{\lambda}(k)}{\lambda} 
\;\mbox{,} \qquad
\qquad f_{\lambda}(k) \,=\,\frac{1}{N_c^2-1}\sum_a 
\,|\Phi^{a}_{\lambda}(k)|^2 \;\mbox{.}
\label{eq:Gk}
\ee
If we consider a FP matrix of the type
${\cal M}^{ab} = - \delta^{ab} \Delta - K^{ab}$
(this is the case in Landau, Coulomb and MAG) then we have
\be
G(k) \, = \,\int_{\lambda_{\min}}^{\lambda_{max}} \frac{d\lambda}{\lambda}\,\rho(\lambda)\,
\frac{1}{N_c^2 - 1}\sum_a
\,|\Phi^{a}_{\lambda}(k)|^2 \;\mbox{,}
\quad
\Phi^{a}_{\lambda}(k) \,=\, \frac{1}{k^2 - \lambda} \, \sum_{x,y} e^{-i k x} K^{ab}_{xy}
\Phi^b_{\lambda,y} \;\mbox{.}
\ee
In a numerical simulation
we look at $G(k_{min})$ when the volume increases
(and $\lambda_{\min}$ decreases). Thus, the IR behavior of $G(k_{min})$
depends on the quantity
\be
\Phi^{a}_{\lambda_{min}}(k_{min}) \,=\, \frac{1}{k_{min}^2 - \lambda_{min}}
\, \sum_{x,y} e^{-i k_{min} x} K^{ab}_{xy}
\Phi^b_{\lambda_{min},y}
\ee
and there is a clear competition between the smallest eigenvalue of the Laplacian 
$\,k_{min}^2 \sim L^{-2}\,$ and the smallest eigenvalue of the FP operator
$\lambda_{min}$.

It is interesting to notice that using Eq.\ (\ref{eq:Gk}) we can easily explain
why finite-size effects are small\footnote{Note that in Ref.\ \cite{Oliveira:2006zg}
there are, actually, strong finite-size effects for the ghost propagator
$G(k)$. However, in that case the effects are probably due to the use of strong
asymmetric lattices, with different ratios of the spatial over the temporal
extension of the lattice.}
when the ghost propagator $G(k)$ is evaluated numerically.
Indeed, it is sufficient to have
$\,\rho(\lambda) f_{\lambda}(k) / \lambda \, \sim  \,
\lambda^{\beta}\,$ with $\beta > -1$ in the limit of small eigenvalues
$\lambda$. In Ref.\ \cite{Sternbeck:2005vs} it has been obtained
(for 4d Landau gauge)
that the quantity $\,R(\overline{\lambda}) \,=\, 
\int_{\lambda_{\min}}^{\overline{\lambda}}
d\lambda\;\rho(\lambda) f_{\lambda}(k) \lambda^{-1} / G(k) \,$
behaves as $\overline{\lambda}^{\nu}$,
with $\nu > 0$, for small $\overline{\lambda}$
considering the two smallest nonzero momenta $k$. This implies
$\beta = \nu - 1 > -1$.


\section{Conclusions}

We believe that the study of the spectral properties of the
FP operator in different gauges can help us understand
the general features of the Gribov-Zwanziger confinement
scenario. In particular, it would be important to
clarify for which gauge conditions the confinement
mechanism can be related to an enhancement of the ghost
propagator in the IR limit.


\begin{theacknowledgments}
The author thanks the organizers for the invitation to present this review at
QCHS7 and A.\ Maas, T.\ Mendes and D.\ Zwanziger for helpful discussions.
This work was partially supported by FAPESP
(under grants \# 00/05047-5 and 06/57316-6) and by CNPq. 
\end{theacknowledgments}


\end{document}